# Superlative spin transport of holes in ultra-thin black phosphorus

Jiawei Liu[1,2], Deyi Fu[1], Tingyu Qu[2], Deqiang Zhang[2], Kenji Watanabe[4], Takashi Taniguchi[4], Ahmet Avsar*[1,3], Barbaros Özyilmaz*[1,2,3]

[1]*Centre for Advanced 2D Materials, National University of Singapore, Singapore 117546,*

[2]*Department of Physics, National University of Singapore, Singapore 117542, Singapore,*

[3]*Department of Materials Science and Engineering, National University of Singapore, Singapore 117575, Singapore*

[4]*National Institute for Materials Science, Tsukuba,*

*Corresponding authors: aavsar@nus.edu.sg, barbaros@nus.edu.sg*

The development of energy-efficient spin-based hybrid devices that can perform functions such as logic, communication, and storage requires the ability to control and transport highly polarized spin currents over long distances in semiconductors. While traditional semiconductors such as silicon support spin transport, the effects of carrier type and concentration on important spin parameters are not well understood due to the need for extrinsic doping, which can cause additional momentum and hence spin scattering. Two-dimensional semiconductors, on the other hand, offer the ability to tune carrier type and concentration through field effect gating and inherently have long intrinsic spin lifetimes, making them a desirable platform for spin transport. Here, we study gate-tunable spin transport across narrow band-gap black phosphorus-based spin valves which enable us to systematically investigate spin transport with varying hole and electron concentrations under non-local geometry. Our findings demonstrate exceptional pure spin transport that approaches intrinsic limit, particularly in the low hole doping range. We achieved record non-local signals reaching 350 Ω and spin lifetimes exceeding 16 ns. Contrary to the behaviour seen in typical semiconductors, we find that the spin transport performance of holes in black phosphorus is significantly better than that of electrons, with the Elliott-Yafet process being the primary spin scattering mechanism. The observation of gate-tunable nanosecond spin lifetimes and colossal pure spin signals in both p- and n-type black phosphorus offers promising prospects for the development of novel semiconducting spintronics devices requiring sharp p-n interfaces.

Spin scattering in materials occurs when an interaction with the environment, such as phonons or impurities, changes the electron spin. The fundamental cause of the main spin relaxation processes, such as Elliot-Yafet (EY)[1,2] and Dyakonov-Perel[3], is the spin-orbit interaction, which results in strongly mixing the spin and orbital momentum eigenstates. Since these two common relaxation processes have a close correlation between spin and momentum scattering events, the electronic properties of materials, which are dependent on carrier type, concentration, and temperature, determine the ability of the material to sustain large spin polarization for a long time. Semiconductors such as Si[4], GaAs[5], and Ge[6] are appealing spintronics material systems as they can enable realizing multifunctional spintronic devices and be easily integrated into transition semiconductor electronics. Their spin transport properties have been extensively studied using the non-local, four-terminal lateral spin valve geometry which has proven to be an effective method in generating and detecting pure spin current[7]. For instance, n-type Si spin valves were shown to support spin injection, transport, and detection with spin lifetimes exceeding nanoseconds even at room temperature (RT)[4]. The weak dependence of spin and charge parameters on gate voltage[8] and temperature[4] was attributed to the EY mechanism, further emphasizing the importance of momentum relaxation on spin scattering. In contrast, the transport and detection of spin-polarized holes in non-local geometry have yet to be demonstrated. For example, while p-type Ge is considered a promising spin channel material due to its higher charge mobility compared to other p-type semiconductors such as Si and GaAs, spin transport measurements could only be demonstrated under the debated three-terminal geometry with spin lifetimes below 1 ps, nearly three orders of magnitude less than those obtained in n-type Ge[9]. There is a need for the identification of p-type materials that exhibit high-quality spin transport (e.g. large spin signals and long spin lifetimes) to be used in multifunctional spintronics devices such as spin filters, unipolar spin diodes, and spin solar cells[10–12], which require the formation of p-n junctions.

The thinness nature of two-dimensional materials uniquely allows for the effective tuning of charge carrier types and concentrations through the application of an electric field, without inducing any disorder. Unlike traditional semiconducting materials, which require chemical doping to enable electron or hole conduction and can negatively impact mobility and spin transport, this gate tunability of 2D materials makes them well-suited for spintronic research and applications as it permits examining the intrinsic spin transport abilities of electrons and holes as a function of carrier concentration[13]. While monolayer graphene exhibits high-quality spin transport in both electrons and holes[14], its semi-metallic nature limits its effective gate tunability, preventing the formation of strong p-n junctions and thus restricting its potential in semiconducting spintronics[13]. This constraint further necessitates the finding of an ideal 2D semiconductor for spintronic research from the diverse 2D materials family.

The observation of high charge mobility and its light atomic mass, leading to weak spin-orbit coupling, has established two-dimensional black phosphorus (BP) as a promising semiconducting material for spintronics research. BP possesses a relatively small but sizeable band gap, enabling ambipolar behaviour with prominent highly conductive electron and hole channels, facilitating the formation of a p-n junction with sharp boundaries. Recent measurements of spin transport in n-type BP under a non-local geometry show gate tunable nanosecond spin lifetimes with large spin signals nearly four orders of magnitude higher compared to what has been measured in Si[15]. The dominant spin scattering mechanism was found to be the EY process, indicating that improving electronic mobilities could lead to even higher quality spin transport properties. Given these intrinsic characteristics, along with the observation of record hole electron mobilities close to 50,000 cm$^2$ V$^{-1}$ s$^{-1}$ [16], make this highly anisotropic material a strong candidate for the first realization of spin-polarized hole transport in a semiconducting material[17].

In this work, we report the observation of superlative hole spin transport in ultra-thin black phosphorus under non-local geometry, which extends to room temperature. For this, we fabricated spin-valves consisting of hBN-encapsulated BP with an optimal thickness of 8 nm for reaching the maximum mobility[18]. The optical images of the representative device before and after the metallization process is shown in Fig. 1b and Fig.1c, respectively. The fabrication process is similar to our previous work on n-type BP spin transport[15], with the exception that we deliberately positioned half of the BP on the hBN$_B$ substrate and the other half on a SiO$_2$/Si wafer, as depicted in Figure 1a. This substrate arrangement together with applied gate voltages allowed us to reach both conduction and valence bands and hence study the influence of charge carrier type and concentration on spin transport in the same flake. The spin characteristics of our three investigated devices are studied as a function of the substrate (hBN$_B$ vs SiO$_2$), the gate voltage (± 80 V), and temperature (1.5 K – 300 K). Throughout the manuscript, we will present results obtained from a single flake for the purpose of direct comparison. Here, sample A represents the spin valve on a SiO$_2$ substrate, while sample B represents the spin valve on an hBN$_B$ substrate.

Prior to any spin transport measurements, the charge transport properties of sample A ($hBN_T$ / BP / $SiO_2$) and sample B ($hBN_T$ / BP / $hBN_B$) were studied using the conventional four-terminal lateral measurement configuration. Figure 1c exhibits the transfer characteristics of these two samples recorded while sweeping the gate voltage at 1.6 K. Both samples display electron and hole conduction, with sample A displaying higher hole conductance and sample B demonstrating electron conductance dominance, despite sharing the same BP crystal as the channel material. The source of this predominant electron conductivity in sample B is its $hBN_B$ substrate, which has been reported to impart additional n-type doping to BP[19]. We note that this substrate-induced doping strategy plays an important role in our study while investigating the spin transport properties of holes and electrons. Here, both samples share the same BP crystal, but our focus will be on the spin transport facilitated by the hole carriers in sample A and the electron carriers in sample B due to their higher current values. The mobility of the charge carriers was extracted based on the field effect transistor, with the mobility of holes in sample A being ~690 $cm^2$ $V^{-1}$ $s^{-1}$ and the mobility of electrons in sample B being ~780 $cm^2$ $V^{-1}$ $s^{-1}$. These similar mobilities of the two samples make it possible to compare the spin transport capabilities of electrons and holes under similar momentum relaxation rates. In addition, the two-terminal I-V characteristics of both devices were studied at different gate voltage values (Fig. 1c - Inset), which show an exponential behaviour in both the hole and electron sides, indicating that the top $hBN_T$ serves as a high-quality tunnel barrier for spin injection and acts as an encapsulation layer.

We now turn our attention to the spin transport measurements under typical non-local spin valve configuration. First, we focus on spin valve measurements in sample A where the charge carriers are holes. For this purpose, a fixed DC bias of |10| µA was applied between the injector and reference electrodes to create a spin accumulation. To detect spin-polarized current, the non-local voltage was recorded using the detector and the second reference electrodes while sweeping a magnetic field parallel to the easy axis of the electrodes in both backward and forward directions between 0.04 T to - 0.04 T. The top panel of Figure 2 displays the spin valve results obtained at $V_{BG}$ = - 60 V (~$n$ = 1.6 x $10^{12}$ $cm^{-2}$) and $T$ = 1.6 K, which exhibits a significant change in the non-local resistances of up to 120 Ω, signifying a marked difference in spin accumulation between parallel and antiparallel magnetization configurations. The spin origin of this signal was verified through conventional Hanle spin precession measurements, shown in the bottom panel of Figure 2, where a magnetic field was applied in the out-of-plane perpendicular direction and the injector and detector electrodes were kept in an antiparallel magnetization configuration. The resulting Hanle curves can be fitted with the solution of the Bloch equation,

$$R_{NL} \propto \int_0^\infty \frac{1}{\sqrt{4\pi D t}} e^{-\frac{L^2}{4Dt}} e^{-\frac{t}{\tau_s}} \cos(\omega_L t) dt$$

where $D$ for the diffusion constant and $L$ for the spacing between the injector and detector, $\tau_s$ for the spin lifetime, $\omega_L$ for the Lamour frequency. We extract a $\tau_s$ value of ~16.1 ns at low temperatures in the hole regime ($n$~1.42×10$^{12}$ cm$^{-2}$). Note that this value measured in p-type BP is comparable to the longest spin lifetimes previously recorded in graphene[14] (~12 ns) and n-type silicon[4] (~16.4 ns) based non-local spin valves. For comparison, the Hanle curves in n-type BP (sample B) under the same carrier concentration was also studied (represented by the black curve in Figure 2, bottom). Here, $\tau_s$ is found to be ~2.2 ns, which is comparable to previous results obtained in n-type BP[15] but nearly one order of magnitude lower than the spin relaxation time we probe in the hole regime even though both samples have comparable momentum relaxation times. These findings are particularly intriguing as theoretical predictions[20] anticipate smaller spin-mixing probabilities for electrons compared to holes in phosphorene, contrasting with our experimental findings.

After establishing clear spin transport in p-type BP, we examine the dependence of spin transport on the hole carrier concentration in sample A. For this, we perform gate-voltage-dependent Hanle precession measurements and extract spin lifetimes at each carrier concentration value. As seen in Fig. 3a, we observe several nanosecond spin lifetimes in all ranges of studied carrier concentration. The spin lifetime values are found to be decreasing as the carrier concentration increases, similar to what we see in the carrier concentration dependence of momentum relaxation times. Here, we note that while sample B also shows a similar gate-voltage dependence trend, the magnitude of spin lifetimes is almost an order of magnitude smaller than those in sample A. In the bottom panel of Figure 3, we then display the ratio of momentum and spin relaxation as a function of carrier concentration, which remains nearly constant throughout, indicating the dominance of EY spin scattering in p-type BP, similar to what has been observed in n-type BP[15]. In this mechanism, the spin relaxation rate is given by

$$1/\tau_s = \alpha\, b^2/\tau_p$$

where $\alpha$ is a perfector between 1 and 4, depending on the details of the impurity or phonon scattering, $\tau_p$ is the momentum relaxation time and $b^2$ is the spin mixing probability which gives the probability to find the electron with a spin up if the state has the average spin pointing down (or vice versa). The $b^2$ values we obtain are remarkable, almost two orders of magnitude better than the value estimated based on calculations of the valance band of phosphorene that only considered spin-orbit coupling of the host lattice without any extrinsic scatterers[15,20]. This result indicates that our device, encapsulated with hBN and fabricated under inert gas environments, has allowed us to realize spin transport near the intrinsic limit.

The high quality of the spin transport of p-type BP is also evidenced by the observation of record non-local spin signals. Figure 3b displays the carrier concentration dependence of the non-local spin resistance. The amplitudes of the spin signal in the hole regime are notably higher than on the electron side, while they follow the same gate dependence. In both regimes, the spin signals increase as the carrier concentration increases, reaching peak values of 320 Ω in sample A and 50 Ω in sample B. The spin polarization values obtained are greater than 40% on the hole side (Fig. 3b-Inset), which is significantly higher than both the inherent spin polarization of Co and the values of $Co/Al_2O_3$ tunnel barriers[21]. To shed light on the importance of these large spin signals, we compare them to those obtained in the best semiconducting and metallic channel materials-based spin valves. As seen in Figure 3c, spin signals in BP are three times larger than in graphene spin valves using the same[22] or other highest-quality[23,24] tunnel barriers and up to four orders of magnitude greater than in traditional materials like Si[4], Ge[6], GaAs[5], Ag[25], Cu[26] and Al[7]. Our observations of gate-tunable nanosecond spin lifetimes and record signals offer promising prospects for developing functional spin-logic devices that rely on large pure spin currents.

We turn our attention to temperature-dependent spin transport measurements. Figure 4 shows the spin relaxation time's temperature dependency, ranging from 1.6 K to RT, at a hole carrier concentration of ~2.5 x $10^{12}$ $cm^{-2}$. This plot demonstrates that the spin lifetime in p-type BP is nearly unaffected by temperature between 1.6 K and 80 K, but drops to half its value when the temperature is increased to RT. Notably, this behaviour mirrors the temperature dependence of momentum relaxation time. By plotting $b^2$ against $T$, it is evident that it remains nearly constant in temperature (as shown in the bottom of Fig. 4), suggesting a dominant EY-type spin relaxation mechanism across the entire temperature range, which is in line with the gate-voltage dependent measurements in Fig. 3a. Here, we conclude that at low temperatures, spin flipping of holes occurs due to spin-orbit interaction from the host lattice and momentum scattering from impurities whereas phonons become the dominant factor in spin relaxation at higher temperatures.

Finally, we discuss spin transport at RT. We observe a clear spin switching and Hanle spin precession with signals surpassing 100 Ω (Fig.4-Inset & Extended Data Fig. 4) and extract a spin lifetime of ~ 3.6 ns and spin diffusion length exceeding 3.1 μm at $V_{BG}$ = - 60 V. This result is significant, as room temperature spin transport in semiconductors under four-terminal non-local geometry is uncommon, with the only exceptions being n-type Si[4] with a spin lifetime of 2.1 ns and a spin signal of a few mΩ, and n-type BP[15] with a spin lifetime of 0.7 ns and a spin signal of 25 Ω. These values could be further enhanced by integrating the p-type BP with the recently discovered ultra-thin nano magnets such as $Fe_3GeTe_2$[27] , due to formed atomically sharp van der Waals interfaces which would reduce spin-dependent scatterings. Such large signals in these bilayers could be also utilized to realize pure spin current-induced magnetization reversal[28] with reduced critical switching current densities at RT. From a technological standpoint, these heterostructures could serve as the building blocks for spin logic devices such as the hybrid metal-semiconductor junctions suggested in *ref* 29.

## ACKNOWLEDGEMENTS

B.Ö. acknowledges support by the National Research Foundation, Prime Minister's Office, Singapore, under its Competitive Research Program (CRP award no. NRF-CRP22-2019-8), the NRF Investigatorship (NRFI award no. NRF-NRFI2018-08), and the Medium-Sized Centre Programme. A. A. acknowledges support by the National Research Foundation, Prime Minister's Office, Singapore. K.W. and T.T. acknowledge support from the Elemental Strategy Initiative conducted by the MEXT, Japan, and the CREST (JPMJCR15F3), JST.

## METHODS

### Measurement

Cryogenic measurements were performed in a closed-cycle Oxford Instruments cryomagnetic system with a base temperature of ~1.6 K. During two-terminal charge transport measurements, drain currents were measured using a Keithley Sourcemeter 6430, and a Keithley Sourcemeter 2400 was used to apply bias through the $SiO_2$ gate dielectric. Four-terminal charge transport measurements were performed by using a lock-in amplifier (Stanford Research SR830) at very low-frequency (~13 Hz) while the device is highly conductive. For gate-dependent spin transport measurements, we utilized a Keithley-2400 current source to apply fixed current and a nanovoltmeter (Keithley 2182A) was used while detecting spin current.

## AUTHOR CONTRIBUTIONS

A.A. and B.Ö. designed and coordinated the work. D. Fu, J. Liu, T. Qu and D. Zhang fabricated the samples. J. Liu performed transport measurements. K.W. and T.T. grew the hBN. J. Liu, A.A. analyzed the results; J. Liu and A.A. wrote the manuscript with input from all authors.

## COMPETING FINANCIAL INTERESTS

The authors declare no competing financial interests.

**FIGURE CAPTIONS**

**Figure 1 | Device geometry & fabrication and charge transport characterization. a,** Schematics of the device. BP is partially deposited on SiO$_2$ (sample A) and hBN$_B$ (sample B) substrates. During spin transport measurements, four-terminal non-local geometry is employed and the electrode pairs of 1&2 and 3&4 are used as spin injector and detector electrodes in sample A and sample B, respectively. Optical image of the device before (**b**) and after (**c**) the metallization process. The thicknesses of the BN substrate (hBN$_B$), BP channel and the encapsulating BN (hBN$_T$) layer are ~20 nm, ~8 nm and ~1 nm, respectively. **d,** Back gate voltage ($V_G$) dependence of conductivity for the BP on SiO$_2$ (blue line) and hBN$_B$ (black line) substrates. BP shows hole (electron) dominated charge transport characteristics on SiO$_2$ (hBN$_B$) substrate. Inset shows the $V_{SD}$ dependences of $I_{SD}$ for both devices at $V_G$ = -60 & 60 V. Measurements were taken at 1.6 K.

**Figure 2 | Carrier type and concentration dependent electronic spin transport and Hanle spin precession measurements.** Top graph: Non-local signal as a function of the in-plane magnetic field in sample A where charge carriers are holes. Yellow and red horizontal arrows represent the sweeping directions of the magnetic field. Bottom graph: Non-local signal as a function of the perpendicular magnetic field. Measurements are performed with an injected current of 10 μA at 1.6 K and carrier concentration is fixed to ~ 1.42 x 10$^{12}$ cm$^{-2}$. The blue line represents measurements taken on sample A (hole conduction). The black line represents the electron spin precession measurements taken in device B (electron conduction). The signal obtained in device B is multiplied eight times for comparison purposes.

**Figure 3 | Carrier type and concentration dependent non-local spin signal.** . **a,** Carrier concentration dependent spin lifetimes extracted from Hanle measurements for hole (blue dots) and electron (black dotes) -types charge carriers. The bottom graph shows the temperature dependence of the ratio of the momentum and spin relaxation times for holes. All of

measurements were taken at 1.6 K. **b,** Carrier concentration and type dependent non-local spin signal obtained in non-local spin valve measurements. Both carriers exhibit similar concentration dependences: the signal decreases after peaking around $1.5 \times 10^{12}$ cm$^{-2}$. The maximum signal obtained in ohms while carriers are hole (sample A) and electron (sample B) are ~ 320 and ~ 50, respectively. Inset shows the hole carrier dependence of spin polarization. **c,** Non-local spin signal as a functional of distance between injector and detector electrodes in highly promising semiconducting and metallic materials-based spin valves. Two-dimensional materials exhibit significantly larger spin signals (light blue region) compared to their three-dimensional counterparts (light red region). For graphene spin valves, the types of tunnel barriers are also indicated.

**Figure 4 | Temperature dependent spin transport.** Temperature dependence of spin and momentum relaxation times at a fixed hole carrier concentration of ~ $2.5 \times 10^{12}$ cm$^{-2}$ in sample A. The bottom graph shows the temperature dependence of the ratio of the momentum and spin relaxation times for holes and electrons. Inset shows the spin valve measurement taken at RT.

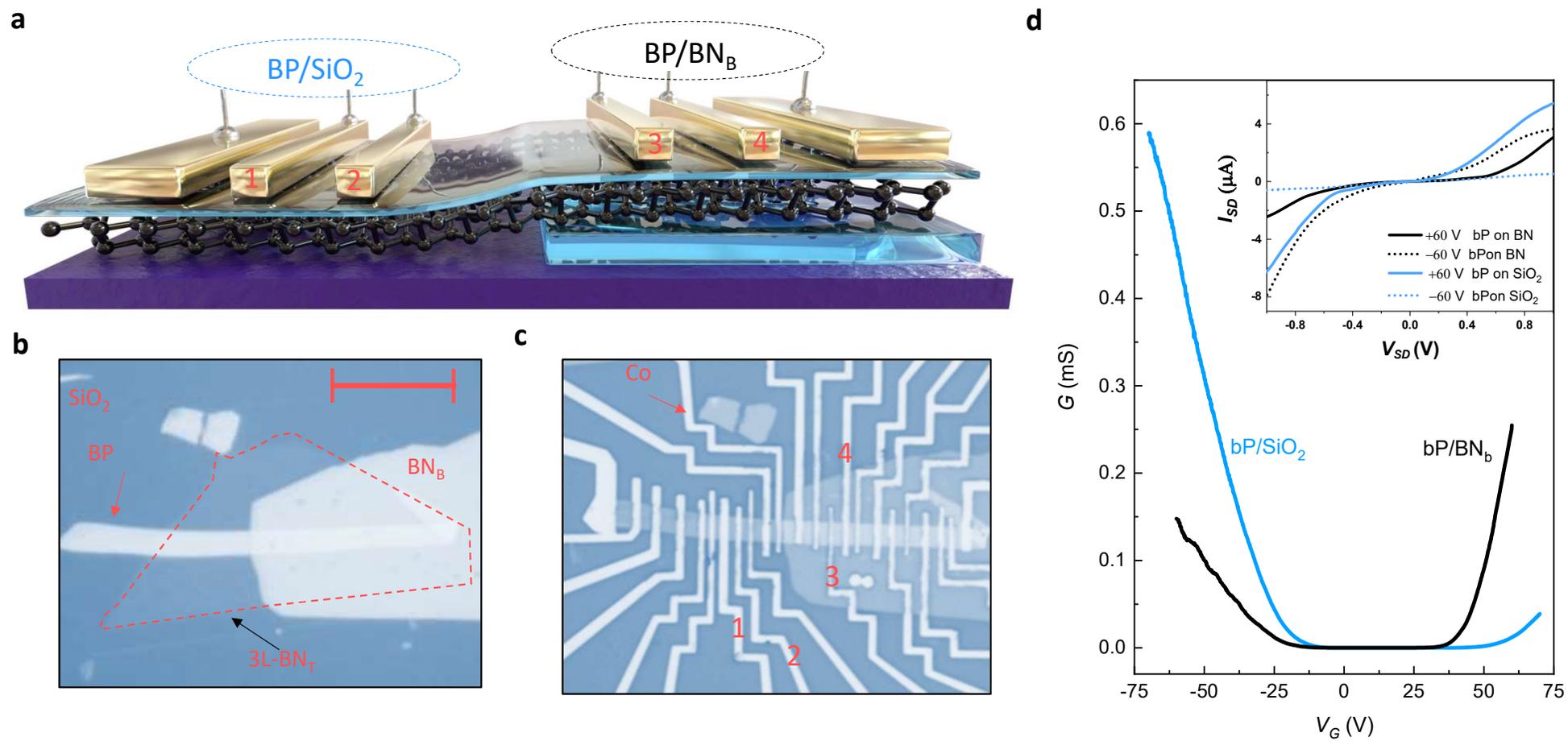

**Figure 1**

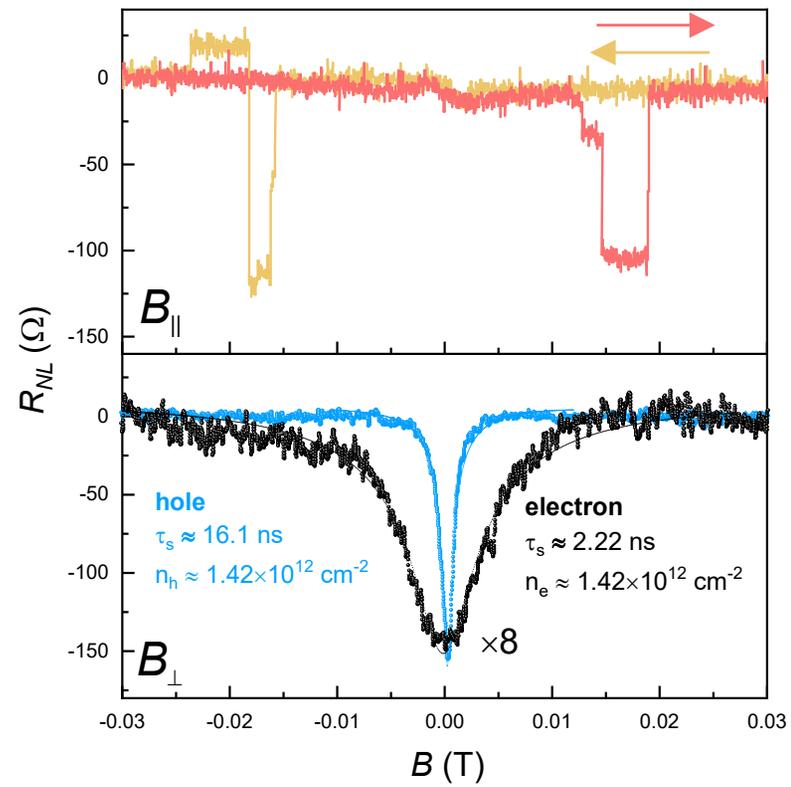

**Figure 2**

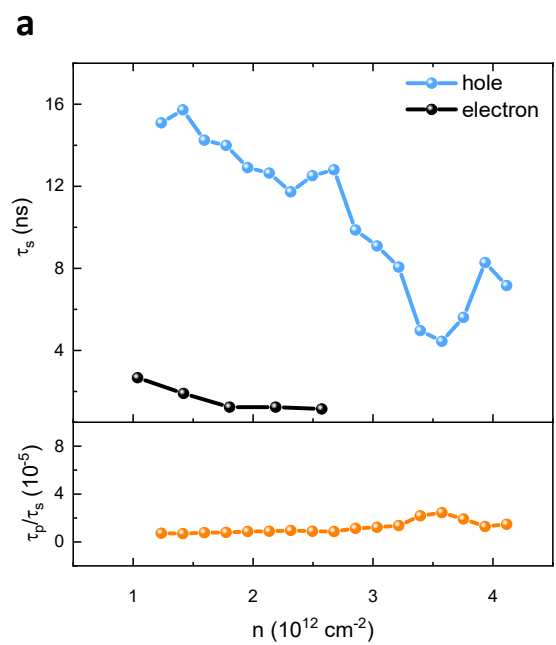 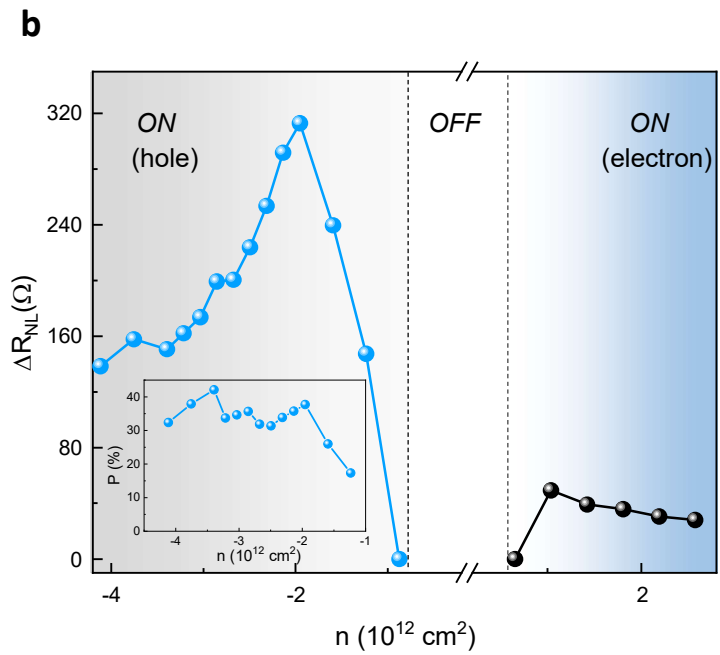 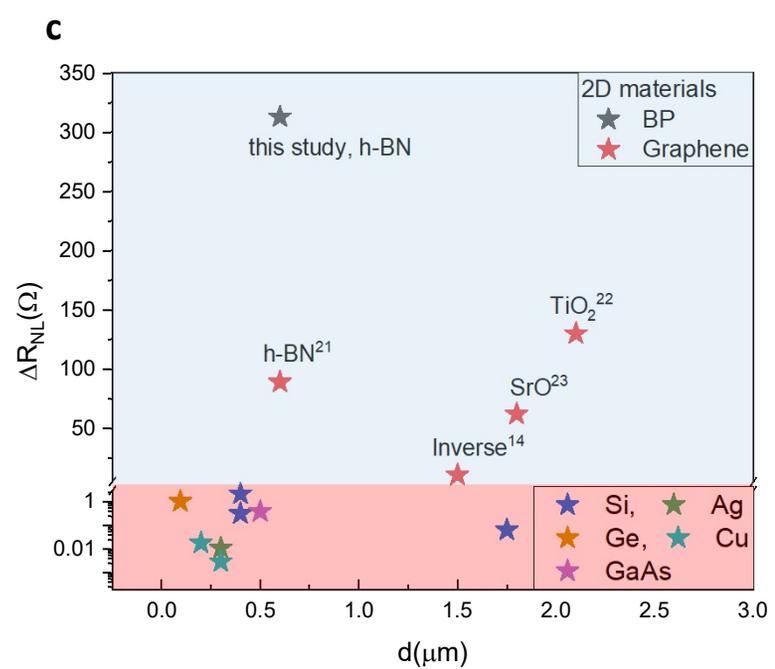

**Figure 3**

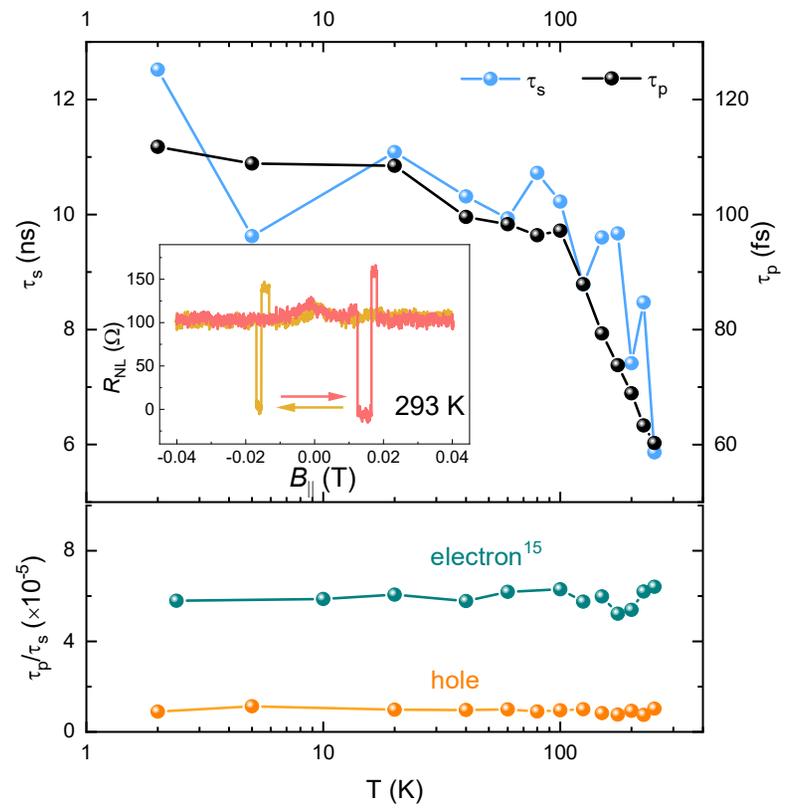

**Figure 4**